\documentclass[aps,pra,twocolumn,superscriptaddress,showpacs]{revtex4} 
\usepackage{graphicx}
\usepackage{amsmath,amssymb,color}

\def\ii{{\mathrm{i}}}
\def\ee{{\mathrm{e}}}
\def\dd{{\mathrm{d}}}

\def\ket#1{|\mbox{$#1$}\rangle}

\def\bracket#1{\langle\mbox{$#1$}\rangle}
\def\bracketi#1#2{\langle\mbox{$#1$}|\mbox{$#2$}\rangle}
\def\bracketii#1#2#3{\langle\mbox{$#1$}|\mbox{$#2$}|\mbox{$#3$}\rangle}

\def\sub#1{_{\scriptsize\mbox{#1}}}
\def\Re{\mathop{\mathrm{Re}}}
\def\Im{\mathop{\mathrm{Im}}}
\def\bigbracket#1{\Bigl\langle\mbox{$#1$}\Bigr\rangle}

\begin{document}

\title{Extracting joint weak values from two-dimensional spatial displacements}

\author{Hirokazu Kobayashi}
\email{kobayashi.hirokazu@kochi-tech.ac.jp}
\affiliation{Department of Electronic and Photonic System Engineering,
Kochi University of Technology, Tosayamada-cho, Kochi, Japan.}
\author{Graciana Puentes}
\email{graciana.puentes@icfo.es}
\affiliation{ICFO---Institut de Ciencies Fotoniques, Mediterranean
Technology Park, 08860 Castelldefels, Barcelona, Spain}
\author{Yutaka Shikano}
 \email{yshikano@ims.ac.jp}
 \affiliation{Institute for Molecular Science, Okazaki, Aichi, Japan.}
 \affiliation{Chapman University, Orange, California, USA.}

\date{\today}

\begin{abstract}
The joint weak value is a counterfactual quantity related to quantum
correlations and quantum dynamics, which can be retrieved via weak
measurements, as initiated by Aharonov and colleagues. In this Rapid
Communication, we provide a full analytical extension of the method
described by Puentes \textit{et al}. [Phys. Rev. Lett. 109, 040401 (2012)], to
extract the joint weak values of single-particle operators from
two-dimensional spatial displacements of Laguerre-Gauss probe states, for
the case of azimuthal index $|l|>1$. This method has a statistical
advantage over previous ones since information about the conjugate
observable, i.e., the momentum displacement of the probe, is not required.
Moreover, we demonstrate that under certain conditions, the joint weak
value can be extracted directly from spatial displacements without any
additional data processing.
\end{abstract}

\maketitle

As the well-known examples of the Einstein-Podolsky-Rosen paradox
and the which-path information in the Young double-slit experiment illustrate, 
quantum dynamics and quantum correlations have
eluded our intuitive understanding for a long time. 
To capture these features, it is important to access the expectation value
of products of two observables since such joint observables can contain
information about quantum correlations and quantum dynamics. 
However, in the standard technique, the strong measurement of joint
observables is difficult to implement because of the requirement of a nonlinear
Hamiltonian~\cite{nonlinear_interaction} 
and the state reduction resulting from the measurement backaction. 
In this regard, \textit{the joint weak value} is a good candidate, as seen in
Refs.~\cite{Resch, PHT, MJP, Hardy}, 
since this quantity can be experimentally
obtained with a tiny measurement backaction from the second-order effects
of the time evolution via a local, linear, and single-particle interaction Hamiltonian.

Historically, the weak value was initiated by Aharonov, Albert,
and Vaidman (AAV)~\cite{AAV}, inspired from the two-time 
formulation of the quantum mechanical system~\cite{ABL}. 
This formulation is characterized by the pre- and post-selected states of the
system. 
When a system initially prepared in state $\ket{\psi\sub{i}}$ 
is post-selected in state $\ket{\psi\sub{f}}$, the weak value of the
observable $\hat{A}$ is defined as 
\begin{align}
\bracket{\hat{A}}\sub{w}&:=
\frac{\bracketii{\psi\sub{f}}{\hat{A}}{\psi\sub{i}}}{\bracketi{\psi\sub{f}}{\psi\sub{i}}},   \label{eq:1}
\end{align}
which can lie outside the spectrum of $\hat{A}$ and can even take an
imaginary number. 
Recently, there have been various experimental realizations using the
weak value within the foundations of quantum mechanics, for example, in the
direct measurement of a wavefunction~\cite{Lundeen}, the which-path
measurement in the Young double slit experiment~\cite{Steinberg}, 
the confirmation of the Heisenberg-Ozawa uncertainty relationship~\cite{ozawa}, the confirmation
of the Hardy paradox~\cite{Hardy_exp}, 
the violation of the Leggett-Garg inequality~\cite{leggett_garg}, 
and the measurement of a geometric phase in an interferometer~\cite{Kobayashi}.

There are also various proposals for extracting the weak value from
experimental data (see for reviews~\cite{AV, AT, Shikano}). 
Originally, AAV restricted their attention to the standard von Neumann
paradigm with a weak interaction Hamiltonian
of the form $\hat{H}=g\hat{A}\otimes\hat{P}_x$, 
where $g$ is a small coupling constant and 
$\hat{P}_x$ is the momentum observable of the probe state conjugate
to the position observable $\hat{X}$ with $[\hat{X},\hat{P}_x]=\ii\hbar$.
Moreover, they assumed that the probe state was initially prepared in a
fundamental Gaussian mode. 
In this case, we can determine the real and imaginary parts of the weak
value from the spatial and momentum displacements of the probe state~\cite{AAV, Jozsa}. 

Recently, Resch and Steinberg~\cite{Resch} proposed a measurement technique for the weak value
of the joint observable as alluded before. 
They employed a two-dimensional Gaussian probe state and a weak coupling
of the system observable $\hat{A}$ ($\hat{B}$) with $x$ ($y$) dimension of the
probe state. 
By performing a second order expansion in the two-dimensional
displacement of the probe state, 
they showed that it is possible to extract the real
part of the joint weak value $\bracket{\hat{A}\hat{B}}\sub{w}$ from the
second-order spatial displacement, under the assumption of commuting
observables $[\hat{A},\hat{B}]=0$.
This procedure, however, needs the real and imaginary parts of single
weak values $\bracket{\hat{A}}\sub{w}$ and $\bracket{\hat{B}}\sub{w}$ to
calculate the joint weak value. 
Thus, we have to obtain full information of the probe wavefunction, i.e., not only the
spatial displacement but also the momentum displacement, by taking the
Fourier transform of the probe wavefunction. 

In this Rapid Communication, we provide a simple method for extracting the joint weak
value only from the spatial displacement of the two-dimensional probe
state. 
This method has a statistical
advantage over previous methods 
since full information of the probe wavefunction is not required. 
The key idea is to employ Laguerre-Gauss (LG) modes for probe states, as was initiated
by Puentes \textit{et al.}~\cite{PHT}.
We extend this idea and provide the full description of the weak measurement for
higher order LG modes with radial index $p = 0$ to extract 
the joint weak values. 

The LG modes are given as the natural solutions of the paraxial
wave-equation~\cite{Siegman} and characterized by a radial index $p$
and an azimuthal index $l$. 
The modes have zero intensity at its center and an annular intensity distribution. 
The wavefront of the LG modes is composed of $|l|$ intertwined helical
wavefronts, with a handedness given by the sign of $l$. 
It has been shown that each photon in the LG modes carries a quantized intrinsic orbital
angular momentum $l\hbar$, in addition to the spin-like angular
momentum $\pm\hbar$ associated with circularly polarized
waves~\cite{PhysRevA.45.8185}. 
The LG modes have been created using various experimental setups, e.g.,  
using spatial light modulators~\cite{Ando} and using the reflection on a conically-shaped mirror~\cite{Kobayashi2}. Furthermore, the LG modes have many 
applications, for example, for achieving high efficiency of optical tweezers~\cite{tweezer}, for reducing the thermal noise inside gravitational-wave 
interferometers~\cite{grav}, and for generating entanglement with high
efficiency~\cite{ent}. 
The amplitude distribution of the LG modes with
radial index $p=0$ is given as
\begin{equation}
	\phi_i (x, y) = N \{ x + i \cdot {\rm sgn} (l) y \}^{|l|} \exp \left( - \frac{x^2 + y^2}{4 \sigma^2} \right),  \label{eq:9}
\end{equation}
where $\sigma$ is the variance in the case of $l = 0$,
$\text{sgn}(\cdot)$ is the sign function, and $N$ is the
normalization constant.
When $l=0$, Eq.~(\ref{eq:9}) corresponds to a fundamental Gaussian
mode. 
In this case, the amplitude distribution is factorable in two
directions, $x$ and $y$. 
When $|l|>0$, however, it is no longer
factorable, and this is a key factor for retrieving the joint weak values. 

Consider a weak interaction between the LG probe state 
$\ket{\phi\sub{i}}=\int\dd x\dd y\phi\sub{i}(x,y)\ket{x,y}$ and an initial state
$\ket{\psi\sub{i}}$ of the system for the
joint weak measurement of the observables $\hat{A}$ and $\hat{B}$. 
The total input state is 
$\ket{\Psi\sub{i}}=\ket{\psi\sub{i}}\otimes\ket{\phi\sub{i}}$. 
For the measurement process, we use the standard von Neumann paradigm with
the local, linear, and single-particle interaction Hamiltonian as follows:
\begin{align}
 \hat{H}&=g\delta(t-t_0)\left(\hat{A}\otimes\hat{P}_x+\hat{B}\otimes\hat{P}_y\right),  \label{eq:2}
\end{align}
where a coupling constant $g$ is sufficiently small, and $\hat{P}_x$ and $\hat{P}_y$ are the
momentum observables of the probe conjugate to two commuting position observables $\hat{X}$ and
$\hat{Y}$, respectively. 
Here we have taken the interaction to be impulsive at time
$t=t_0$ and the same coupling constant $g$ between $x$ and $y$
directions for simplicity. 

After the interaction between the system and probe states, we post-select the system
in state $\ket{\psi\sub{f}}$, resulting in the probe state:
\begin{align}
\ket{\phi\sub{f}}&=
\bracketii{\psi\sub{f}}{\ee^{-\ii g(\hat{A}\otimes\hat{P}_x+\hat{B}\otimes\hat{P}_y)/\hbar}}{\psi\sub{i}}
\ket{\phi\sub{i}}.  \label{eq:3}
\end{align}
We denote the expectation value of an observable $\hat{M}$ in the initial probe state
$\ket{\phi\sub{i}}$ and the final probe state $\ket{\phi\sub{f}}$ as 
\begin{align}
\bracket{\hat{M}}\sub{i}\equiv\frac{\bracketii{\phi\sub{i}}{\hat{M}}{\phi\sub{i}}}
{\bracketi{\phi\sub{i}}{\phi\sub{i}}}
,\hspace*{0.1cm}
\bracket{\hat{M}}\sub{f}\equiv\frac{\bracketii{\phi\sub{f}}{\hat{M}}{\phi\sub{f}}}
{\bracketi{\phi\sub{f}}{\phi\sub{f}}}.  \label{eq:4}
\end{align}
We can calculate the displacement $\bracket{\hat{M}}\sub{fi}$ by expanding the time
evolution operator in Eq.~(\ref{eq:3}) up to the second order in the coupling constant $g$:
\begin{align}
\bracket{\hat{M}}\sub{fi}&\equiv\bracket{\hat{M}}\sub{f}-\bracket{\hat{M}}\sub{i}\nonumber\\
&\simeq\frac{2g}{\hbar}\Im\bracket{\hat{M}\hat{H}_1}\sub{i}
+\frac{g^2}{\hbar^2}\left(
\bracket{\hat{H}_1^\dagger\hat{M}\hat{H}_1}\sub{i}-\Re\bracket{\hat{M}\hat{H}_2}\sub{i}
\right),  \label{eq:6}
\end{align}
where
\begin{align}
\hat{H}_1&\equiv
\hat{P}_x\bracket{\hat{A}}\sub{w}+\hat{P}_y\bracket{\hat{B}}\sub{w},    \label{eq:7}\\
\hat{H}_2&\equiv\hat{P}_x^2\bracket{\hat{A}^2}\sub{w}+\hat{P}_y^2\bracket{\hat{B}^2}\sub{w}
+\hat{P}_x\hat{P}_y\bracket{\hat{A}\hat{B}+\hat{B}\hat{A}}\sub{w}.  \label{eq:8}
\end{align}
$\hat{H}_1$ and $\hat{H}_2$ are respectively the first- and second-order terms of the
time evolution operator in the coupling constant $g$. 
In case of the large weak value, the approximation in
Eq.~(\ref{eq:6}) is invalid because of the measurement
backaction~\cite{backaction}. 
In this Rapid Communication, however, we treat the weak measurement as a 
powerful tool for addressing the counterintuitive quantum phenomena seen in
Refs.~\cite{Lundeen,Steinberg,ozawa,Hardy_exp,leggett_garg}. 
Thus, the
weak value is predetermined, and it corresponds to the so-called fixed
value. 
In this case, the measurement backaction can be ignored 
by setting a sufficiently small coupling constant. 

First, we calculate the spatial displacement of the probe state in the 
$x$-$y$ plane. In this case, we employ Eq.~(\ref{eq:6}) up to the first
order and obtain the spatial displacements as
\begin{align}
 \bracket{\hat{X}}\sub{fi}&=g\left(\Re\bracket{\hat{A}}\sub{w}+l\Im\bracket{\hat{B}}\sub{w}\right),  \label{eq:10}\\
 \bracket{\hat{Y}}\sub{fi}&=g\left(\Re\bracket{\hat{B}}\sub{w}-l\Im\bracket{\hat{A}}\sub{w}\right). \label{eq:11}
\end{align}
These results show that the spatial displacements along the $x$ and $y$
directions include not only the real part of the weak values but also its
imaginary part except for $l=0$. 
This is because the probe state given in Eq.~(\ref{eq:9})
is not factorable in the $x$ and $y$ dimensions so the coupling of the
observable $\hat{A}$ ($\hat{B}$) with the $x$ ($y$) dimension of the probe also affects its
$y$ ($x$) dimension as the imaginary part of the weak value. 

Next, we consider the joint measurement of two position observables along the
mutually-perpendicular directions in the probe state.
The joint observables in the probe state can be represented 
by $\hat{X}\hat{Y}$ and $\hat{X}^2-\hat{Y}^2$. 
The other products of the perpendicular position operators can be represented by 
the linear superposition of these two observables. 
The expectation values of these joint observables can be experimentally
obtained from the two-dimensional intensity
distribution captured by the imaging sensor. 

We employ Eq.~(\ref{eq:6}) up
to the second order in the coupling constant $g$ and obtain the second-order spatial displacements as
\begin{widetext}
\begin{align}
 \bracket{\hat{X}\hat{Y}}\sub{fi}&=g^2\Bigl[
-\frac{l^2-|l|-2}{4}\Re\left(\bracket{\hat{A}}\sub{w}\bracket{\hat{B}}\sub{w}^*\right)
+\frac{l^2-|l|+2}{4}\Re\bigbracket{\displaystyle\frac{\hat{A}\hat{B}+\hat{B}\hat{A}}{2}}\sub{w}
-l\Im\bigbracket{\displaystyle\frac{\hat{A}^2-\hat{B}^2}{2}}\sub{w}
\Bigr],  \label{eq:12}\\
 \bigbracket{\displaystyle\frac{\hat{X}^2-\hat{Y}^2}{2}}\sub{fi}&=
g^2\Bigl[
-\frac{l^2-|l|-2}{8}\left(|\bracket{\hat{A}}\sub{w}|^2-|\bracket{\hat{B}}\sub{w}|^2\right)
+\frac{l^2-|l|+2}{4}\Re\bigbracket{\displaystyle\frac{\hat{A}^2-\hat{B}^2}{2}}\sub{w}
+l\Im\bigbracket{\displaystyle\frac{\hat{A}\hat{B}+\hat{B}\hat{A}}{2}}\sub{w}
\Bigr].  \label{eq:13}
\end{align}
\end{widetext}
For the fundamental Gaussian probe state ($l=0$), 
our result is consistent with the result
given in Ref.~\cite{Resch}. 
In this case, Eqs.~(\ref{eq:12}) and  (\ref{eq:13}) contain only the real part of the
joint weak values $\bracket{\hat{A}\hat{B}+\hat{B}\hat{A}}\sub{w}$ and
$\bracket{\hat{A}^2-\hat{B}^2}\sub{w}$. 
For $|l|>0$, however, we obtain additional terms proportional to their
imaginary part as illustrated in Ref.~\cite{PHT} for the
case of $|l|=1$~\cite{erratum}.
Thus, the second-order spatial displacements contain all the information about 
joint weak values. 
If there are a sufficient number of different outcomes for the first- and
second-order spatial displacements, 
we can calculate the joint weak values from
Eqs.~(\ref{eq:10})--~(\ref{eq:13})~\cite{noncommutative}. 
This feature is not attainable with a fundamental Gaussian mode
($l=0$). 

One useful method for obtaining the joint weak values is to take two
separate measurements by using two probe states with different $l$
values. 
From a practical point of view, the best choice is two probe states with 
equal magnitude but different signs of $l$ since they can be easily
prepared by using a mirror reflection. 
In this method, the two separate measurements 
with four types of spatial displacements bring eight outcomes, 
which equals to the number of unknown real and imaginary parts of single and joint weak values. 
Thus, it is possible to calculate the real and imaginary parts of
the joint weak values from
Eqs.~(\ref{eq:10})--~(\ref{eq:13}). 
The advantage of this method is that we can use the same interaction
Hamiltonian, i.e., the same experimental setup; we only need to change the input probe state. 

In particular, in case of $|l|=2$, the coefficients of the first-order weak values
  $\bracket{\hat{A}}\sub{w}$ and $\bracket{\hat{B}}\sub{w}$ in
  Eqs.~(\ref{eq:12}) and (\ref{eq:13}) are eliminated. Therefore, only
  the joint weak values remain. 
By using the two probe states with $l=\pm 2$, we can extract the joint
  weak values only from the second-order spatial displacement as follows:
\begin{align}
\Re\bracket{\hat{A}\hat{B}+\hat{B}\hat{A}}\sub{w}
&=\frac{\bracket{\hat{X}\hat{Y}}_++\bracket{\hat{X}\hat{Y}}_-}{g^2},\\
\Im\bracket{\hat{A}\hat{B}+\hat{B}\hat{A}}\sub{w}
&=\frac{\bracket{\hat{X}^2-\hat{Y}^2}_+-\bracket{\hat{X}^2-\hat{Y}^2}_-}{4g^2},\\
\Re\bracket{\hat{A}^2-\hat{B}^2}\sub{w}
&=\frac{\bracket{\hat{X}^2-\hat{Y}^2}_++\bracket{\hat{X}^2-\hat{Y}^2}_-}{2g^2},\\
\Im\bracket{\hat{A}^2-\hat{B}^2}\sub{w}
&=-\frac{\bracket{\hat{X}\hat{Y}}_+-\bracket{\hat{X}\hat{Y}}_-}{2g^2},
\end{align}
where $\bracket{\hspace*{0.1cm}\cdot\hspace*{0.1cm}}_+$ and 
$\bracket{\hspace*{0.1cm}\cdot\hspace*{0.1cm}}_-$ correspond to the
spatial displacements for the $l=+2$ and $-2$ cases, respectively. 

Moreover, under the assumption of $\hat{A}^2=\hat{B}^2$, the joint weak
  value $\bracket{\hat{A}^2-\hat{B}^2}\sub{w}$ is also eliminated. 
Thus, we can derive a simple and direct relationship
  between the second-order spatial displacements and the joint weak value as 
\begin{align}
\Re\bracket{\hat{A}\hat{B}+\hat{B}\hat{A}}\sub{w}
&=\frac{2}{g^2}\bracket{\hat{X}\hat{Y}}\sub{fi},  \label{eq:17}\\
\Im\bracket{\hat{A}\hat{B}+\hat{B}\hat{A}}\sub{w}
&=\frac{\text{sgn}(l)}{2g^2}\bracket{\hat{X}^2-\hat{Y}^2}\sub{fi}.  \label{eq:18}
\end{align}
With these equations, the joint weak value can be directly extracted 
from the spatial displacements by using a single LG probe state with $|l|=2$. 
Although our consideration is restricted only to the case of 
$\hat{A}^2=\hat{B}^2$, this case includes many experimental setups. 
For example, the joint weak measurement of Pauli operators
$\hat{\sigma}_i$ $(i=x,y,z)$ on different
two-level systems, such as $\hat{A}=\hat{\sigma}_z\otimes\hat{I}$ and
$\hat{B}=\hat{I}\otimes\hat{\sigma}_z$ with the identity operator
$\hat{I}$, 
is included since they satisfy
the property $\hat{A}^2=\hat{B}^2=\hat{I}$. 

In this Rapid Communication, we derive the full description of the weak measurement
for the LG probe state and provide a simple method for extracting the joint weak
values only from the spatial displacements of the two-dimensional probe
state. Our method has a statistical advantage over previous ones
since information about the momentum displacement is not
required. Moreover, by using the LG probe state with $|l|=2$, 
and for the case $\hat{A}^2=\hat{B}^2$, 
we can extract the full joint weak value, i.e., the real and
imaginary parts of the joint weak value directly from the
second-order spatial displacements without any additional data
processing. 

In our study, we restrict the probe state to the LG mode with the radial
index $p=0$. 
However, we can use other spatial modes, e.g. LG modes with
nonzero radial index $p$, Hermite Gauss modes, and hypergeometric-Gaussian
modes. 
There remains an interesting problem on the relationship between the
spatial rotational symmetry of the probe state and the joint weak
value since the spatial rotational symmetry in our restricted case is
broken after the weak measurement.
Moreover, it would be interesting to study the case of probe states
given by superpositions of different spatial modes.
By using such a extended probe state, the condition
$\hat{A}^2=\hat{B}^2$ might be relaxed. 

We thank Y. Aharonov, J. Tollaksen, A. M. Steinberg, and L. Vaidman for useful comments and
suggestions. 
One of the authors (GP) acknowledges financial support from the Marie Curie Incoming Fellowship COFUND.

\bibliographystyle{junsrt}

\begin{thebibliography}{99}
\bibitem{noncommutative}
	It is noted that one can compute $\bracket{\hat{A}\hat{B}+\hat{B}\hat{A}}\sub{w}$
        but not $\bracket{\hat{A}\hat{B}}\sub{w}$ or
	$\bracket{\hat{B}\hat{A}}\sub{w}$ 
	except for $[\hat{A}, \hat{B}]=c \hat{I}$, where $c$ is the
        imaginary number and $\hat{I}$ is the identity operator.
\end{thebibliography}

\end{document}